# Astronomical x-ray polarimetry based on photoelectric effect with microgap detectors.

*Paolo Soffitta[a,*], Enrico Costa[a], Giuseppe di Persio[a], Ennio Morelli[a], Alda Rubini[a]*

*Ronaldo Bellazzini[b], Alessandro Brez[b], Renzo Raffo[b], Gloria Spandre[b],*

*David Joy[c]*

[a]Istituto di Astrofisica Spaziale del CNR, Via Fosso del Cavaliere 00133, Rome, Italy

[b]INFN- Pisa, Via Livornese 582, I-56010 S. Piero a Grado, Pisa, Italy

[c] Science and Engineering Research Facility, University of Tennessee, Knoxville, TN37996-0810

## *Abstract*

*The polarisation of x-ray photons can be determined by measuring the direction of emission of a K-shell photoelectron. Effective exploitation of this effect below 10 keV would allow development of a highly sensitive x-ray polarimeter dedicated in particular to x-ray astronomy observations. Only with the advent of finely segmented gas detectors was it possible to detect polarisation sensitivity based on the photoelectric effect in this energy range. Simulation and measurements at 5.4 and 8.04 keV with a microgap gas counter, using both a polarised and an unpolarised x-ray source, showed that the photoelectron track in a neon-based gas mixture retains the memory of the polarisation of the incoming photons. Possible experiments aimed at galactic/extragalactic sources and solar flares are considered and their sensitivity to these sources is calculated.*

---

[*] Corresponding author. Tel. +39-06-4993-4006; fax: + 39-06-20660188;e-mail:soffitta@ias.rm.cnr.it

## 1. Introduction

Our study of the photoelectric effect in a gas is aimed at developing a polarimeter for x-ray astrophysics. We have based our discussion on astrophysical applications, although the results achieved could be of more general interest.

In the x-ray band, Bragg diffraction at 45° and Compton scattering around 90° are the two basic techniques for measuring linear polarisation. Both of them are applied in the Stellar X-Ray Polarimeter (SXRP) ([1],[2]), the only astronomical mission with large throughput optics. In the framework of these techniques, a major advance with polarimeters based on a combination of detectors and analysers at the focal plane is not feasible. It would also be unrealistic at present to design another mission based on long focal ratio optics. Hence, significant progress in astronomical x-ray polarimetry could only be made by adopting different techniques.

## 2. Measurement concept and basic formalism

A polarimeter is usually composed of an analyser and a detector, which are rotated around the pointing axis. During rotation the counts are modulated with the law:

$$C(\phi) = A + B \cdot \cos^2(\phi - \phi_0) \ , \ (1)$$

where $\phi$ is the (azimuthal) angle between the projection of the polarisation vector on the analyser and the characteristic axis of the polarimeter and $\phi_o$ is the phase angle.

So, the counts span from $C_{min} = A$ to a $C_{max} = A+B$. The important parameter of the polarimeter for a 100 % polarized source is usually referred to as the *modulation factor* and it is defined as:

$$\mu = \frac{C_{max} - C_{min}}{C_{max} + C_{min}} \quad . \quad (2)$$

In a 'non dispersive' polarimeter, such as a Compton polarimeter, photons analysed at different phases are simultaneously detected, but the formalism is basically the same. In all this formalism we use relations between expectation values, while in reality actual counts are added to a background and distributed around expectation values according to Poisson statistics. When the source and background rates follow Poisson statistics, the sensitivity of a polarimeter like the one just described is defined by the minimum modulated flux needed to exceed, at the desired level of confidence, the expected statistical fluctuation both of the background and of the unpolarised fraction of the source. This is the so-called minimum detectable polarisation (MDP) and, for a given source, in a net observing time T, at a level of $n_\sigma$ standard deviations (in Poisson distribution), is [3]:

$$MDP(n_\sigma) = \frac{1}{\varepsilon\mu} \cdot \frac{n_\sigma}{S} \cdot \sqrt{2\frac{\varepsilon S + B}{AT}} \quad , (3)$$

where $S$ is the source flux, $\varepsilon$ the detector efficiency, $\mu$ the modulation factor, $A$ is the collecting area and $B$ is the background counting rate per unit surface. For slightly different formulations we prefer to factorise the efficiency so that we can separate the factor of merit in the most common case of observation dominated by the background counts ($B \gg \varepsilon S$):

$$Q = \varepsilon\mu / \sqrt{B} \quad , (4)$$

which is typical of a given technology or configuration. Once the technology is assessed, the surface of the instrument and the observation time can be determined to plan a certain measurement

Another very interesting feature is the capability of keeping systematic errors under control. Some of the candidate astrophysical targets for an x-ray polarimetry programme are expected to show polarisation in "a few %" range. Hence, any polarimeter must control (namely the capability to prevent or to post facto correct) systematic effects to the level of 1% or, possibly two or three times better. The limit flux detectable from the polarimeter is fixed by these two main parameters.

In other branches of x-ray astronomy, gas-filled or solid-state detectors based on photoelectric absorption have in the past provided sufficient efficiency and background rejection (excellent in imaging systems). In x-ray polarimetry this achievement has been impossible as no significant modulation with polarisation has yet been obtained with photoelectric devices and the two techniques (Bragg and Compton) used so far are overwhelmed by low efficiency and high background, respectively. Attempts to carry out photoelectric polarimetry have been unsuccessful because of the low modulation or the high systematic effects [4], [5]. In this paper we show the results of testing very high resolution detectors as photoelectric polarimeters.

### 3. Photoelectric effect and related polarimetric capabilities

In 1926, Auger [7], by irradiating a cloud chamber with polarised x-rays discovered, together with the relevance of dielectronic recombination (Auger effect), the dependence of the direction of created photoelectrons on the linear polarisation of the original photons. A simplified quantum-mechanical treatment such as the one in Heitler provides a distribution [8] that is valid in the nonrelativistic regime:

$$\frac{\partial \sigma}{\partial \Omega} = r_o^2 \frac{Z^5}{137^4} \left(\frac{mc^2}{h\nu}\right)^{7/2} \frac{4\sqrt{2} \sin^2(\theta) \cos^2(\varphi)}{(1-\beta\cos(\theta))^4}, \qquad (5)$$

where θ is the angle between the direction of the incoming photon and the ejected electrons, while φ is the azimuth angle of the latter with respect to the x-ray polarisation vector, Z is the atomic number of the photoabsorbing atom, $r_o$ is the classical radius of the electron β is the photoelectron velocity in fraction of c. The term in the denominator accounts for a slight bending forward of the distribution with the photoelectron energy. For example, 8-keV electrons are ejected maximally at about 75$^o$ with respect to the incoming photon direction. Due to the angular dependence on φ, the photoelectric effect is 100% modulated by x-ray polarisation for whatever θ (while in the Compton effect, sensitivity to polarisation decreases for scattering angles that are not 90°). Thus, this process is potentially a perfect analyser for a polarimeter.

Starting from this point many workers have tried to conceive or develop polarimeters based on the distribution of (5). The main limitation to this technique is that at the energies of interest electrons propagate in the matter less than photons. Furthermore, they scatter on atoms and the original direction of the track is randomised while the photoelectron energy decreases.

Elastic electron scattering from interaction with nuclei is mainly responsible for the randomisation. Elastic scattering is described by the angular differential form [cm$^2$ sterad$^{-1}$] of the screened Rutherford cross section [9]:

$$\frac{d\sigma}{d\Omega} = 5.21 x 10^{-21} \frac{Z^2}{E^2} \left( \frac{E+511}{E+1024} \right)^2 \frac{1}{\left( sin^2\left(\frac{\phi}{2}\right) + \alpha \right)^2} \quad (6)$$

where ϕ is the deflection angle with respect to the incoming direction and α is the 'screening factor'[10], E is the electron energy [keV]. If the scattering angle is large, as in the case of low-energy electrons (<20 keV), the Rutherford formula is no longer accurate and must be replaced by the Mott cross section. For this we use a parameterised analytical form [11]. Below 1 keV, it is necessary to use numerically tabulated differential cross-section data [12], [13].

The electron energy decreasing due to bremsstrahlung or to discrete inelastic scattering can be modelled in the Bethe formula [14]:

$$\frac{dE}{dS} = -78500 * \frac{Z}{AE} \log_e \left( \frac{1.166E}{J} \right) \quad (7)$$

where J is the mean ionisation potential [15] and A is the atomic weight, E is the electron energy [keV] and S is the product of the density ρ of the material of the medium [g/cm$^3$] and the distance travelled along the trajectory s. For very low energy the Bethe expression (7) implies an unphysical change of sign. In this energy regime the stopping power increases with energy [16]. The Joy and Luo equation [17] takes into account both regimes and agrees with experimental data. Inelastic collisions with orbital electrons produce deflections of the order of ΔE/E rad, hence negligible in the keV regime ([18] and reference therein). To prevent elastic scattering from destroying the polarimetric information, researchers used a very thin target. If the photon impinges on the target at grazing angle, the absorption efficiency can be made high, the escape probability for the photoelectron can also be reasonably high and the electron can be detected in a microchannel plate. Since this escape probability depends on the direction of the photoelectron, unless randomised by scattering, the pair grazing target/electron detector acts as a polarimeter. This design has been very popular since some measurements [19] suggested that low-energy electrons emitted from a cathode could retain more memory of the polarisation than expected from current theories of collision. Further measurements have significantly reduced this effect [5]. Some measurements have also been made with thin targets by collecting energetic photoelectrons with high electric fields. It is clear that some polarimetry can be done using this approach, but with a few major limitations, including the loss of information on the energy, poor efficiency and extreme sensitivity to systematic effects.

A different approach is to have a multipixel detector and observe the "cross-talk" of contiguous pixels. When the electron is produced close to the edge of one pixel it will enter the nearby pixel with a probability strongly dependent on the direction and hence on the polarisation. This will result in a pair of coincident signals from contiguous pixels. This method has been attempted [4] by injecting polarised photons into a multiwire proportional counter in a parallel direction to the wires by looking for coincident signals from nearby cells.

More recently, charge-coupled devices (CCDs) [20], [21], [22], [38], [39], [40] have been tested with polarised x-rays and simulated. At energy below 10 keV optics for X-ray astronomy are very effective. In this energy range the actual pixel size of a CCD are still larger then the photoelectron track length and single pixel events outnumbers two pixels events, called pairs, and moreover multi-pixels events for a 'true' electron tracking. In this energy range therefore this approach is still not very effective, as only the few photons absorbed close to the pixel edge produce pairs whenever an electron leaks from one pixel to the adjacent. The orientation of these pairs is modulated with the polarisation angle. However those events are extremely sensitive to systematic effects, when the pixels are not uniformly illuminated. This is because the probability of a photoelectron switching on two contiguous pixels could depend more on the probability of being absorbed in the frontier zone at a particular phase than on the polarisation itself. Those effects can be controlled and calibrated reliably in laboratory [38]. The control and calibration of those effects are much more difficult in an astronomical observation because of the satellite pointing instability and because the Point Spread Function of the optics can be of the same size of the CCD pixels.

A good photoelectric polarimeter should produce an ionisation pattern that depends as much as possible on the polarisation and as little as possible on the absorption point. This can be achieved if the pixel size is much smaller than the photoelectron track length. At energies below 15 keV, where the grazing incidence optics is highly efficient and the source fluxes are high, this condition can now be achieved with gas detectors. A polarimeter based on photoabsorption in a gas with a pixel size much smaller than the photoelectron track could benefit from the good imaging and background rejection capabilities of photoelectric detectors, plus the polarimetric capability provided by high-resolution imaging of the track.

Historically, 30 years ago, Sanford [6], who tried to relate the rise-time of a xenon-filled counter with the length and orientation of the projected photoelectron track, performed the first approach of this type. Sanford rotated the counter by 90$^o$ with respect to the polarisation of the beam (16.5 keV) and got two slightly different distributions of risetime with a small amount (about 2%) of faster signals in the case of electric vector parallel to the anode. This limited result discouraged him from continuing but we can now say that different choices in terms of filling gas and structure of the electric field could lead to more interesting results. In fact, the choice of xenon is not very suitable because we know that the photoelectron distribution is less modulated with the polarisation if the photoelectron comes from L or M shells [23] and elastic scattering is high in xenon. Moreover, the detector used for this test had a cylindrical geometry, which significantly reduced the information on the original shape of the electron cloud because of relevant diffusion along the drift direction.

Recently, the rise time in a Xe-filled proportional was measured by means of a waveform digitiser [24]. The authors found that the risetime of the digitised signals was significantly different if produced by two beams with orthogonal polarisation states and, also, that such a difference increased with energy. These results should be interpreted with some caution at least around the K-edge of Xenon, as pointed out by the same authors, since the modulation measured at 40 keV is larger with respect to the one found right below the K-edge, while the corresponding L-shell photoelectron right below the K-edge have much larger range with respect to the K-shell photoelectron at 40 keV.

Another approach was followed by Austin [25]. He reported a modulation measurement of 30% for 60-keV photons by using an x-ray imager filled with argon. The choice of such a gas establishes a threshold in energy estimated by the authors to be around 30 keV, above which it is possible to sample the size of the charge cloud itself. This is mainly due to the reduced range of the photoelectrons [26] and to the long drift region. Recently, a different kind of photoelectron imager, more oriented to lower energy photons, was proposed as an x-ray polarimeter [27].

The aim of our experiment is to find a modulation at energies below 10 keV, where the grazing incidence optics can focus x-rays.

## 4. Experiment

We have chosen the relatively recent microgap detector [28] instead of other more established configurations for the following reasons:

i) The microgap detector is more suitable than a multiwire proportional chamber, which is based on induction in several cathodes set on a different plane from the anodes, with the related large smearing effects. Also, it is very difficult to build a wire proportional counter with a pitch smaller than 1 mm due to electrostatic and spark problems. A much higher spatial resolution can be reached after measurement of the centre of gravity of the track from averaging of induced signals and no fine structure below the wire pitch can be really appreciated.

ii) The microgap detector is more suitable than a drift or a time projection chamber because it is self-triggered and no independent device that triggers the multiplication process and the acquisition is requested to evaluate the $t_0$. Such a characteristic is essential for an experiment in x-ray astronomy.

iii) The microgap can have a smaller pixel size than the gas microstrip.

## 4.1 Detector

We used a 1-D microgap detector with 128 10-µm-wide anodes with a pitch of 200 µm, and eight cathodes parallel to the anodes (see Fig. 1). Each cathode thus faces a group of 16 anodes and, in our application, is used to trigger the conversion of the signals from the anodes. This 1-D system is not the ideal solution, but it is suitable for deriving, by rotating the device, the main parameters of the polarimeter. This cathode geometry is useful for providing a trigger based on the total energy detected and has low sensitivity to systematic effects.

The entrance window was 25 µm of Mylar aluminised on one side. The absorption/drift gap, defined as the distance between the window and the microgap plate, was 3 mm. In the usual design for these detectors the drift field is defined by three voltages - anode, cathode and grounded window. No field-forming rings or grids are included.

Such a thin absorption gap would be a bad feature for a real astrophysical polarimeter because of the low efficiency. We decided to perform measurements with such a small gap in order to separate the effects of the gas due to primary photoelectron scattering from the effects of diffusion during the drift to the sensing plane. After finding a good configuration, efficiency can be increased by increasing the gap thickness, taking into account the major diffusion of the track during the drift.

The detector is operated typically at gains of 2000, which is reached with voltages suitable for the gas-mixtures we used.

### *4.2. Source*

Polarised photons are obtained by scattering the photons produced by an x-ray generator (25 kV, 500-µm beryllium window, Cu or Cr anode) at 90°. The scatterer is a lithium target (6 mm in diameter, 70 mm long) canned in a beryllium case (500 µm thick) in order to prevent oxidation and nitridation from air [29]. A double-diaphragm collimator between the tube and the scatterer and a multihole collimator at the output limit scattering angles to $(90 \pm 5)^o$ so that radiation impinging on the detector is linearly polarised better then 95%. The photons exit from the collimator at a few hundred per second, which is sufficient for the experiment. The source/scatterer ensemble can be mechanically interfaced with the detectors by a frame. The plane of polarisation can be changed by rotating the detector on the plane perpendicular to the exit direction of the scattered photons by means of a rotary table.

Unpolarised photons are obtained from a $Fe^{55}$ radioactive source, so we have 5.4- and 8.0-keV polarised photons (with some contamination from bremsstrahlung photons limited as much as possible by the pulse height selection) and 5.9-keV unpolarised photons.

The collimators and the lithium scatterer presented to the x-ray source form part of a more complex system that makes use of a smaller x-ray (2 W anodes) tube and a rotary table to assemble a small, portable, polarised x-ray source [30].

### *4.3. Electronics*

The electronics (see Fig.2) is based on the concept of independent analysis and A/D conversion of single strips. Each channel is preamplified and amplified. Threshold discriminators produce digital signals while the analog signals are delayed and fed into a multiplexed ADC. If the decision of the logic circuitry is positive (mainly if there is a signal on one or more cathodes), the ADC is gated and the output of the 128 channels is converted.

The result of the conversion is a set of 128 numbers giving the pulse height for each ADC input at the time of the trigger. We call it a record. It is transferred via CAMAC bus to a computer and recorded in a memory location.

This circuitry is mainly designed to exploit the fast response of microgap detectors and may be not optimal from the point of view of noise and energy resolution. For an astrophysical application with a limited counting rate and more requirements on energy resolution, slower electronic chains could provide better performance. Nevertheless, the present setup is suitable for demonstrating the capabilities and detecting the problems of the proposed method. The noise is different for different channels and may vary with time or a different setup of the experiment. Thus, we collected a rich sample of "noise" events. The trigger is randomly generated to collect those noise events, as well as the records of "good" x-ray events (triggered by the presence signal on the cathodes). These data are used to set thresholds, *fix* the efficiency in low-level track detection and exclude spurious tracks that could arise from random noise on contiguous bins.

## 4.4 Gas mixture

According to (4), the choice of the filling mixture can have an impact on sensitivity through the efficiency, the modulation factor and the background counting rate. Our choice of neon as primary gas for x-ray polarimeters was based on such parameters, as explained below.

*-The efficiency ($\varepsilon$).* This is mainly a matter of absorption, so higher Z and thicker absorption gaps are in principle better. However, for observations in which the source dominates the background, the decrease in sensitivity, for lower Z gas, is effected only through $\sqrt{\varepsilon}$.

*-The modulation factor($\mu$).* The original distribution of $\sin^2 \theta \cdot \cos^2 \varphi$ is smeared and isotropized by several (i-iv) effects, all decreasing the modulation factor. Such a decrement is less visible in low-Z gases.

i) The p, d electrons are less modulated with polarisation than s electrons [23]. A high modulation can only be achieved by operating above the K absorption edge in the gas.

ii) The Auger electron is isotropic; if it bears a significant fraction of the original energy, modulation is smeared. In low-Z gases (organic, Ne, A) the polarimeter should be operated at energies at last twice the K absorption edge. In higher Z gases (Kr, Xe) K photoabsorption followed by fluorescence, if identified by an independent stage or device, could also be used.

iii) The photoelectron scatters several times while slowing down. The original direction is lost and the end of the track looks like a skein (Fig. 3). Residual modulation (if any) is due to the first part of the track, straighter and with a low ionisation density. Modulation increases with energy. Tracks of electrons of the same energy in different mixtures are more or less straight according to the ratio stopping power/scattering. This ratio decreases with Z.

iv) The electron cloud following ionisation by the photoelectron diffuses in the gas during the drift. Modulation will therefore depend on the drift distance. Absorption gaps must not be too thick and the mixture must include heavy complex molecules that reduce the diffusion.

- *The background (B).* Empirically, the background is lower with a lower Z filling gas.

Previous measurement of x-ray polarisation by means of devices able to measure the size of the cloud charge were carried out using a detector filled with gas with a relatively high atomic number. Following the above discussion, we believe that the best trade-off design is obtained with mixtures based on as low Z as possible, compatible with the band of the measurement.

For comparison, the ranges for three noble gases are shown in Fig. 4. Also the probability of elastic scattering is higher when gases with higher atomic number are used. To reach a sensitive modulation below 10 keV, it is thus necessary to fill a microgap proportional counter with mixtures based on neon or helium for which the range can be significantly higher than the anode pitch available. Actually, a better compromise between modulation factor and efficiency could be explored by using denser mixtures based on polyatomic gas for which the probability for an electron to be slowed down by ionising collisions should be sensibly higher than the probability to be deviated by scattering. It is worth stressing here that an optimal mixture or another energy range should also be matched with an optimal detector pitch. The pitch selected is particularly suitable for Ne-based mixtures.

## 4.5 Data analysis

We analysed data to measure the response of a microgap gas proportional counter to a polarised x-ray source. We found that there is a significant difference in the response for polarisation vectors perpendicular or parallel to the direction of the strips. The anode strip pattern collects the tracks via regular and discrete sampling of the charge cloud. In this way each strip produces a signal proportional to the collected charge, which depends on the local charge density. Each event consists of a sequence of amplitudes, one for each fired strip, and the total amplitude of the cluster is proportional to the photon energy.

Care was taken not to add noise events. We sampled the r.m.s. noise for each channel before acquisition and we fixed a threshold according to the maximum r.m.s. noise measured. We considered only the channels for which the amplitude, after pedestal subtraction, exceeded the fixed threshold of *2.5* times the maximum r.m.s. noise. The size measurement should be independent of the particular location of the x-ray interaction, so we chose a single threshold for all the acquisition above the noise level of the noisiest pixels. Also, as the track is divided into small portions, a too high threshold could result in neglecting parts of the track with a low *dE/dS*. This is particularly true for the initial part of the track, which carries the most information on the original direction of the photoelectron.

We define the 'size' of a cluster as being the total number of contiguous strips containing the centre of gravity. This requirement decreases the probability of picking up a noise event. We measured the "mean asymmetry" of all the charge clouds produced by polarised x-ray photons by the parameters of the size distribution and in particular by the mean size. In the case of a (few) non-working channels, we interpolated the signal amplitude of the two adjacent channels. Therefore, we were able not only to increase the statistic but also to get information on the mean size of the distribution along the whole active detector surface. In the following, we describe in detail the results obtained by using a mixture based on neon. Other measurements were performed by using mixtures based on argon and helium. In the helium-based mixture, the track length was too long with respect to the dimension of the detector with a consequent loss of charge. In the argon-based mixture, the detector pitch did not sample the charge adequately.

## 4.6 Results with polarised x-ray source

The most significant set of data was collected with a mixture of Ne (80%) dimethylether (20%). A first rough analysis of these measurements can be found in Soffitta et al. [31,32]. A small sample of events accepted by the selection criteria is shown in Fig 5. As the entire track is projected onto 4 to 14 strips their main features are clear. Some events showed on one side a small peak, which is consistent with the energy released by an Auger electron of 800 eV for an 8-keV photon. If successfully located, the true position of an x-ray interaction can be evaluated by searching for the Auger peaks. But at this stage we give up any ambition of pattern recognition and look for modulation of some very robust estimator of the original orientation of the track. Hence, we neglect any use of the pulse height of individual terms of the track (only a window on the total pulse height is set) and call "track size", the number of contiguous pixels belonging to a single event, as discussed in the previous section. The distribution of these track lengths should be different for different polarisation angles. The three graphs in Fig. 6 show the cluster size distribution for a 5.4-keV polarised source (left), a 5.9-keV unpolarised $Fe^{55}$ source (middle) and an 8.04-keV polarised source. Each one of these three graphs refers to areas of the detectors that are homogeneous in performance. The size distribution for the polarised sources, shows different mean values and shapes for the two planes of polarisation. No difference is shown by the data derived from the $Fe^{55}$ source in the two experimental setups. This excludes systematic effects in the measurements with polarised source, which could derive from a different density or chemical composition in the mixture or a different electronic noise. As expected the difference in shape of the distribution for different planes of polarisation is more evident for 8-keV copper photons. This is as expected due to the increased probability of elastic scattering with decreasing photoelectron kinetic energy, as stated by the Rutherford formula.

Fine detector segmentation and effective sampling of the primary charge cloud allows us to check whether our measurements are consistent with the semiempirical formulae for the extrapolated range [26]. We perform the checking by further manipulation of the cluster size distribution (shown in Fig. 6) to build a "Bragg curve" and measure from the data the extrapolated range at different energies. We can build a Bragg curve from the data if we consider that each strip represents a thin gas foil and we can build "foils" by adding strips. We can get a Bragg transmission curve if we count the number of events whose size is greater than a given number of strips. Figure 7 reports (left) the procedure and the linear fit to the extrapolated range. In the same figure we present (right) comparison between the extrapolated range measured at the three energies and the expected range [26]. The good agreement demonstrates that no major losses in the charge cloud are present in the data after setting the threshold as also discussed in 5.2.

The good local performance of the microgap detector as a polarimeter is shown in Fig. 8. We grouped the centre of gravity in five strips and for these events we measured the mean size. We repeated this measurement for each set of five strips and obtained the results shown in the figure. Thus the total detector area, including the regions containing the strips with interpolated amplitude, was spanned. The error bars are evaluated by dividing the r.m.s. of the size distribution by the square root of the number of counts.

Table 1 contains the complete results of analysis of the measurements in terms of mean cluster size $\lambda$, modulation factor $\mu$, efficiency $\varepsilon$, and counts necessary to detect the measured difference in mean cluster size ($\lambda_\perp$ and $\lambda_{//}$).

### 4.7 Measurement with unpolarised x-ray source

The capability to retain the memory of the initial polarisation can also be verified experimentally by studying the charge spatial distribution of the tracks produced by unpolarised photons, which impinge on a single interaction point. Actually, if the single track does not lose the memory of the initial photoelectron direction, the co-ordinates of the centre of gravity of each cluster will be distributed around the interaction point in a "donut" shape.

A "perfect" donut in the centre of gravity distribution comes from application of Eq. (7) in the case of slowing down without scattering, because in the first part of the track the stopping power is smaller with respect to the final interactions where the energy becomes smaller.

If the scattering is completely dominant, the centre of gravity will always be coincident with the interaction point.

We performed a test by placing a 36-cm-long collimator with an exit hole of 50 μm in front of the unpolarised copper anode x-ray tube. Figure 9 (top) shows the space distribution of the centre of gravity of the tracks. If the tracks are not completely randomised a relation should exist between the centre of gravity position and the cluster sizes as the range is fixed. The longer the cluster size, the more the centre of gravity position should be shifted with respect to the interaction point. We measured [Fig. 9 (bottom)] this effect by selecting the events with a cluster size greater than five channels and we found the bimodal shape expected as per the donut distribution if complete randomisation does not occur.

No quantitative information on the capability of a microgap as a polarimeter is easily derivable from an unpolarised x-ray source. However, by fixing the interaction point and analysing the pattern of the barycentre position, further confidence in the positive result of the measurement with the polarised source was obtained.

## *4.8 Modulation factor for microgap detectors and polarimeters not based on counting*

In order to predict the real performance of the polarimeter, it is necessary to convert the difference in the distribution of the cluster size into useful quantities such as the modulation factor, as usually defined for a polarimeter based on the count rate difference. In this way the definition of the minimum detectable polarisation already stated in (3) could be properly used, and compared with other, already established devices. Such a relation cannot be based only on the value of the mean of the two distributions. The width of the cluster size distribution and the total number of counts must also be included. We could thus define a modulation contrast in analogy to the case of a polarimeter based on the risetime of signals from proportional counters [24]:

$$\mathrm{mod\,contrast} = \frac{\lambda_\perp - \lambda_{//}}{\sigma} \qquad 8$$

where $\lambda_\perp$ represents the mean size for polarisation normal to the strip and $\lambda_{//}$ the mean size for polarisation parallel to the strip and:

$$\sigma = \sqrt{\frac{(\sigma_\perp^2 + \sigma_{//}^2)}{2}} \qquad 9$$

However, we think that a more intuitive definition can be stated as follows.

If we can think of a microgap as a polarimeter, which means that it is sensitive enough to measure the asymmetry of a charge cloud, after a proper normalisation of the size distributions obtained by two measurements with X-ray having orthogonal polarisation states, the following can be stated:
For polarisation normal to the strip direction, the events with asymmetry along this direction represent what, in a polarimeter based on a counting rate, are the counts detected when the electric field is parallel to the polarimeter axis. The events with asymmetry at $90^0$ are, instead, similar to the counts detected when the electric field in the same device is rotated $90^0$. The first events represent the "$C_{max}$", while the latter represent the "$C_{min}$" in formula (2). The same concept applies, but with reversed definition, when the polarisation is along the strips of the microgap. We can measure the asymmetry for our events by setting a borderline in a number of strips (or in another parameter value). Regarding this borderline we recognise these two families of events. Let us again take the measurement with the polarisation vector normal to the direction of the strips themselves. By setting a borderline for a given number of strips, we consider all the events that are longer than the threshold as being oriented along the electric field of the detected photons. If, instead, we consider the measurements with the polarisation vector parallel to the strip direction, the events with asymmetry along the electric field are shorter than the same threshold. In this way, we can sum in "phase" the two classes of events for the two acquisitions as follows and (2) can be rewritten as:

$$\mu = \frac{(N_\perp L + N_{//} S) - (N_\perp S + N_{//} L)}{(N_\perp L + N_{//} S) + (N_\perp S + N_{//} L)}$$

where $N_\perp L$ are the number of events that are longer (L) than the threshold for polarisation normal ($\perp$) to the strip, while $N_{//} S$ are the number of events smaller (S) than the threshold for polarisation parallel (//) to the strip. A corresponding definition applies for $N_{//} L$ and $N_\perp L$.

Therefore, for each borderline we have a polarimeter with its own modulation factor. We group events according to the borderline that gives the maximum modulation, which is, in fact, the modulation factor for a microgap proportional counter. By simply propagating (10), we can also assign a statistical error to the measurement of the modulation factor.

Figure 10 shows a typical curve (for the 8-keV copper line source) for the modulation factor, obtained by changing the value of the borderline. The curve tends to zero at both ends. It is interesting to note that if we neglect in the analysis all the events included in the strip close to the borderline, the modulation is increased because the tracks close to the borderline are not very sensitive to polarisation. Of course, also the efficiency is reduced because of the photons dropped out but the MDP for faint sources can decrease. Actually, from the data, the modulation factor for 8.04 keV copper line increases to 20.5 % by neglecting counts with a cluster size between 4 and 6 and the efficiency is a factor 0.59 lower. Since for bright sources MDP scales as $(\mu\sqrt{\varepsilon})^{-1}$ this selection provides a 5 % better sensitivity. For faint sources MDP scales as $(\mu\varepsilon)^{-1}$, therefore, depending on the source strength and spectrum, it is possible to fine-tuning, via software, the selection of the event size and therefore the choice of the borderline for obtaining the best sensitivity.

## *5. Simulation*

In the following we present the results of the Monte Carlo simulation we performed not only to explain some of the experimental data obtained but also to have a valuable tool for further investigation on the polarimetric capability of devices based on the photoelectric effect. The good agreement between our experimental results and the simulation is impressive.

### *5.1 Monte Carlo programme and source list.*

We carried out our Monte Carlo simulation by using the SS_MOTT code described in [18]. The code performs a single scattering simulation using the Mott cross section in the analytical parameterised form given by [11], which provides an accurate model of the angular deflections experienced by low-energy electrons. The accuracy of this Monte Carlo model is further enhanced by the use of the improved electron stopping power formula of [17], which calculates the energy lost between two subsequent elastic scattering events.

To the original Pascal code we added the possibility to input a list of parameters from an electron

"source" file including the electron starting energy, position, and directions, and the production of a "result" file containing the positions and energy of each elastic scattering event simulated by the SS_MOTT programme.

The source file provides mono-energetic electrons whose angular distribution follows (5) for the fraction of polarised x-ray photons, while the remaining unpolarised fraction follows (5) integrated on φ.

In the reference frame of SS_MOTT, photon polarisation is directed along the y-axis, while the photon that generates the electron is directed along the z-axis.

Interactive Data Language (IDL) is used throughout to create the source electron list and to analyse the result file list. We reproduced a cluster by grouping the energy lost for each track in a 200-µm bin and, starting from a set of simulated clusters, we performed the same analysis as applied to the experimental data.

## *5.2 Comparison between simulation and experimental data*

From a single Monte Carlo run we can derive the expected modulation factor. We actually obtain the parameters described in the formula (1) simultaneously because we produce a simultaneous cluster size distribution for the two orthogonal polarisation states. The difference in the simulated cluster size distribution for two polarization states at 8 keV (Fig. 11) is very similar to the difference found in the data (Fig. 6) obtained at the same energy. We ran the Monte Carlo for different input energies and found that the two modulation factors at 8.05 and 5.4 keV are in very good agreement with the simulation as shown in Fig.12. Such good agreement and the good agreement in the measured and calculated extrapolated range (Fig.7), suggest that no major losses in the charge cloud are present in the data after setting the threshold. In order to have a quantitative evaluation of effect we computed the modulation factor for different thresholds. We found that the modulation factor at 8.04 keV decreases from 11.0 % to 9 % when the threshold increases of 40 % and increases to 12 % when the threshold is decreased of 40 %.

Diffusion is negligible in our data since the drift region is only 3 mm thick. Data and calculation of diffusion [33] for similar mixtures show dependence both on the gas mixture and on the electric field. To evaluate the effect of diffusion in our measurements, we obtained the diffusion coefficient at a drift-field of 7 kV/cm for our mixture based on 80% Ne and 20% Dimethyl-ether (DME) by interpolating calculations at the same drift-field for a mixture 50% Ne and 50% DME and for pure DME. The diffusion coefficient in our case is 220 μm√L(cm) for a maximum of 120 μm in the 3-mm gap of our experimental setup and an average of 85 μm, thus smaller than the pixel size of our detector. The extrapolated range produced by a photoelectron in the mixture we used is about 1100 μm at 5.4 keV and 2000 μm at 8.04 keV; thus, the effect of diffusion is at level of about 10% or less. The good agreement between experimental and simulated data is further evidence that the effect of diffusion is negligible.

Figures 13.1 and 13.2 show the simulation of the measurement of the collimated unpolarised source at 8.04 keV when all the events are collected and only the events with a cluster size larger than five are considered. For these events the loci of centroids have a bimodal distribution, indicating that the scatter does not randomise the track, so the memory of the initial photoelectron direction is conserved even if attenuated by the electron scattering.

## *5.3 Expected sensitivity.*

We evaluated the sensitivity of the microgap polarimeter, considering two possible experiments. In one experiment a microgap detector is placed at the focus of a SODART optics [34], an 8-m focal length telescope to be flown aboard the Spectrum-X Gamma satellite. Such an experiment allows direct comparison with SXRP, which is to date the only x-ray polarimetry experiment dedicated to celestial nonsolar x-ray sources. A drift region of 6 cm is foreseen for the same neon–DME mixture with a drift field of 2 kV/cm where the minimum in the transverse diffusion occurs [33]. At this drift field, the diffusion is 110 $\mu m\sqrt{L(cm)}$, thus the average diffusion which scales with the square root of the distance is still smaller than the pixel size (200 $\mu$m) of the detector. The modulation factors calculated by the simulation, not including diffusion, do not, therefore, sensibly deteriorate. In contrast to the Bragg stage, and also to the Thomson stage of SXRP, a microgap polarimeter is not background-limited at low source flux, as the particle background is negligible because of the neon-based mixture [35]. Also, a microgap rejects charged particle background very efficiently due to segmentation of the anode. We can reach a MDP of 11.5 % for a source of 5 mCrab in $10^5$ s, which is the standard observing time for x-ray polarimeters in astronomy, in the 4-12 keV band. The band is optimised by taking into account the energy dependence of the modulation factor and the detector efficiency. If our 1-dimensional microgap detector is placed in rotation at the focus of the SODART telescope, its MDP is, at this flux level, doubly better than the Bragg stage at 2.6 keV and seven times better than the Thomson stage, which operates in the same energy band. Also, since the track length of the photoelectron is by far larger than the pixel size of the detector, possible misalignment and angular drift between the source direction and satellite pointing would affect the measure of linear polarization much less with respect to the Bragg and Thonson stages in the SXRP experiment. At 1 mCrab the microgap polarimeter MDP is 25.7%, while both the Bragg and the Thomson stage of SXRP are not sensitive at this flux level.

For an experiment using a microgap to study solar flares with the microgap, we could design three 5x5 $cm^2$ detectors with strip directions oriented 30$^o$ apart for a simultaneous measurement of polarisation during the impulsive phase with any optics. The number of independent chains would be 750, which could be read and processed with chips designed as an Application Specific Integrated Circuit. A MDP of 1% can be reached for solar flares in the X class and 2% in the M class in 10 s and in the 4-12 keV energy range. The relatively good spectral capability of the microgap detector allows estimation of the contribution of thermal x-ray emission in this energy band.

## *6. Conclusion*

We measured a positive sensitivity of microgap neon-filled gas detectors to x-ray polarisation below 10 keV. In this energy band, x-ray optics are very effective in imaging celestial sources, thus a highly sensitive polarimeter could be developed. We found a modulation factor of $(11 \pm 2)$ % at 8.04 keV and $(7.2 \pm 2.9)$ % at 5.4 keV. We simulated the propagation of photoelectrons generated by a polarised x-ray beam at different energies and obtained consistent results.

To our knowledge this is the first time that sensitivity to x-ray polarisation has been proved below 10 keV using detectors with a pitch smaller than the photoelectron track length.
Such segmentation gives enormous advantages in the development of x-ray polarimeters. The primary charge cloud can be effectively sampled on a photon-by-photon base. Since the cloud is produced by a K-shell photoelectron, the modulation is, initially, 100 %. We can then build up statistics on parameters measured in the event cluster. We can choose the best parameters, i.e. those that produce the highest sensitivity. We measured a difference in the mean of the cluster size distribution for two orthogonal polarisation angles and used the different shapes of the cluster size distribution to evaluate the modulation factor of the polarimeter. Nevertheless, this is a rough and indirect estimation of the direction of photoelectron emission. More complete exploitation of the experimental information could be conceived in order to increase the sensitivity. In particular application, to increase the sensitivity, also gas mixture based with higher Z element, which anyhow have higher electron scattering, can be studied. The major efficiency of the gas can, in this case, compensate the lower polarization sensitivity of the p photoelectron (60 %, [23]) and the 'blurring' effect of the Auger electron.

We also noted that the response of a polarimeter based on finely segmented detectors is practically insensitive to the interaction point, so an x-ray polarimeter experiment on board a satellite will have reduced requirements in pointing accuracy and stability compared to the standard techniques.

We used a microgap in a configuration typical for a particle beam experiment, hence a gas thickness of 3 mm. In this condition, we can neglect the contribution of diffusion of the primary charge to the sensitive plane, but we foresee that even 6 cm can be used without losing polarisation sensitivity. The average efficiency between 4 and 12 keV is, in this case, 17 %. Better sensitivity compared to SXRP is reached at the focus of a SODART optics, while effects resulting from inclined x-ray beams are small. High sensitivity to solar flares is reached in a fairly small experiment with any optics. Polarisation sensitivity can also be improved by choosing a polyatomic high-density gas mixture with more stopping power in order to further reduce the elastic scattering contribution. Also the contribution of the Auger electrons, which reduce the sensitivity at low energy, is smaller compared to a neon-based mixture. A microgap with a finer pitch is thus required. In this regard, a big step forward has been made in developing bi-dimensional gas detectors and pixel gas detectors such as the microdot [36]. Microdots, at present constructed with a pitch of 100-200 μm [37], can be considered the final goal for gas-detector-based x-ray polarimeters used as focal plane instruments for x-ray astronomy. Using a pixel gas with a "true" pixel readout, it is possible to reach a very high modulation factor. Actually, algorithms that estimate the direction of the photoelectron track can be derived as has been shown for CCD polarimeters [39]. In addition, once the interaction point has been detected by means of pattern recognition, direct measurement of the photoelectron emission angle would be possible, with a dramatic increase in sensitivity. Such a device would allow simultaneous measurement of position, energy spectrum and energy-resolved linear polarisation for all the sources in the field of view.

## 7. Acknowledgements

This research is funded by Agenzia Spaziale Italiana (ASI). We thank the referee for carefully reading the text and making suggestions and useful comments.

# Figure captions

Figure 1. Sketch of the microgap used in the experiment.

Figure 2. Electronics setup for the microgap gas chamber.

Figure 3. Simulated tracks produced by a 4.6-keV photoelectron in neon gas as derived from a 100% polarised source. The photon is assumed to travel along the Z direction, while the polarisation vector is directed along the Y direction. The two top and the bottom-left graphs represent the tracks projected onto the three Cartesian planes. The fourth graph represents the total track length evaluated by the simulation.

Figure 4. Comparison between electron ranges below 10 keV in three different noble gases as derived from Iskef [26].

Figure 5. Sample of four clusters derived from polarised 8.04 keV x-ray photons. Two clusters show a secondary peak whose amplitude is consistent with that produced by an 870-eV Auger electron. Microgap detectors can have better position accuracy if Auger electrons can be identified in the cluster.

Figure 6. Cluster size distribution for a 5.4-keV polarised source (left), 5.9-kev unpolarised $Fe^{55}$ source (middle) and 8.04-keV polarised source (left).
Each graph refers to areas of the detectors that are homogeneous in performance. Sensitive dependence on polarisation is evident only for the polarised source. The measurement with an unpolarised $Fe^{55}$ source shows that no major systematic effects are present in the two orthogonal setups.

Figure 7. Bragg curve (left) evaluated by the cluster size distributions of fig. 6 and evaluation of the extrapolated range. Comparison between measurement and estimation is shown on the right.

Figure 8. Mean of the cluster size distribution after grouping the events with the centroid within 5 microgap strips. The graph shows the good local performance of the microgap.

Figure 9.
Distribution in space of the barycentre of each track measured by an unpolarised 8-keV x-ray source (top). Effect on the distribution in space of the barycentre when tracks with longer strip size are selected (bottom). The bimodal shape expected from the "donut" distribution shows that complete randomisation does not occur. The asymmetry in the peak counts (bottom) are probably due to local non homogeneity in the detector/readout system.

Figure 10. Modulation factor curve (obtained for the 8-keV copper line source) produced by changing the borderline as explained in § 4.8. The modulation factor is represented by the maximum of the curve.

Figure 11 Simulated cluster size distribution for the two polarisation states at 8.04 keV.

Figure 12. Modulation curve obtained by simulation for different energies and comparison with experimental data at 5.4 and 8.04 keV

Figures 13.1 and 13.2. Simulation of the collimated unpolarised source at 8.04 keV when all the events are collected

Figure 13.2 Simulation considering all the events with a cluster size larger than 5. Both simulation and measurement show that the photoelectron does not randomise.

| Energy (keV) | $\lambda_{//}$ | $\lambda_{\perp}$ | $\mu(\%)$ | cts | $\varepsilon(\%)$ |
|---|---|---|---|---|---|
| 8.04 | 4.89±0.03 | 5.31±0.02 | 11.0±2.0 | 160 | 0.6 |
| 5.4 | 3.71±0.04 | 3.90±0.01 | 7.2±2.9 | 370 | 1.9 |
| 5.9 | 4.00±0.02 | 4.01±0.01 | 0.7±3.0 | - | - |

Table 1. Summary of experimental results with polarised and unpolarised x-ray sources. Mean of the cluster size distribution for the two polarisation states with respect to the strip direction ($\lambda_{//}$ and $\lambda_{\perp}$) for the energy used. Modulation factor $\mu(\%)$. Estimated number of counts to detect the measured difference in mean cluster size at the confidence level of 0.1 % (**cts**). Efficiency of absorption $\varepsilon(\%)$

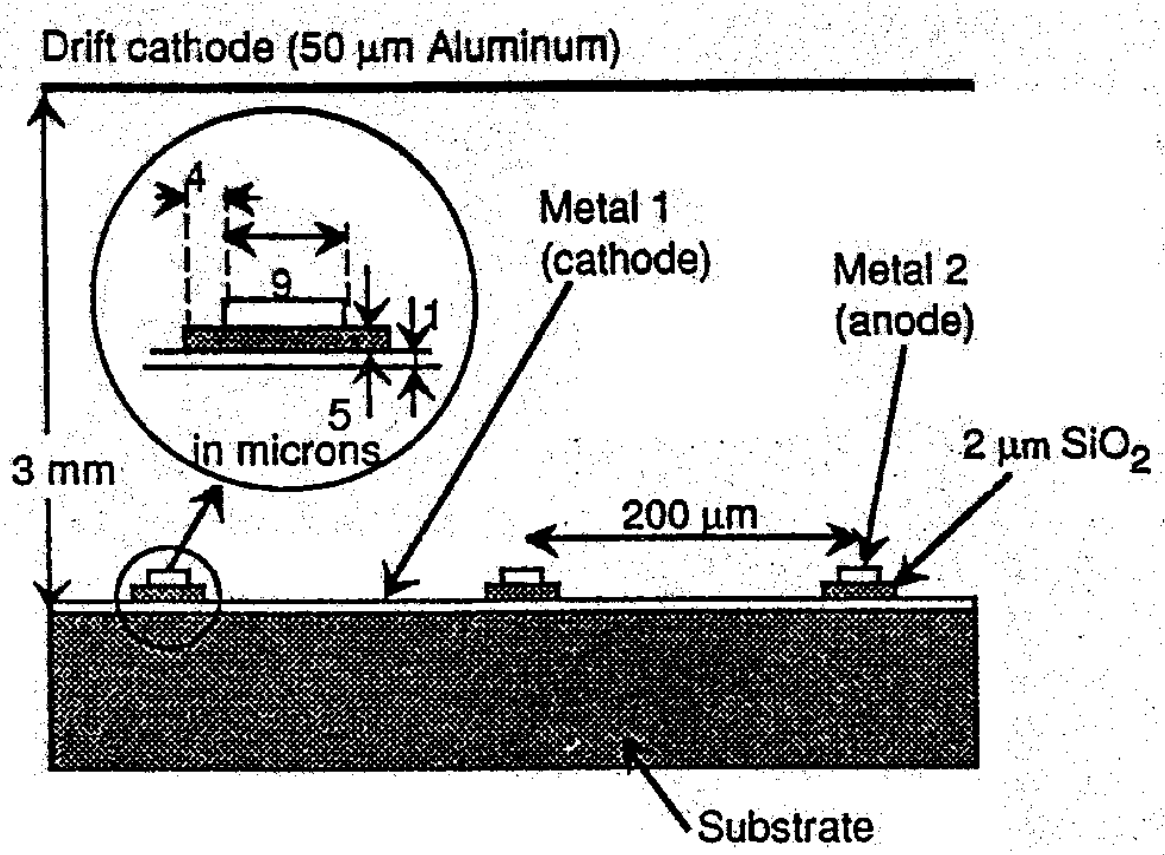

Fig. 1

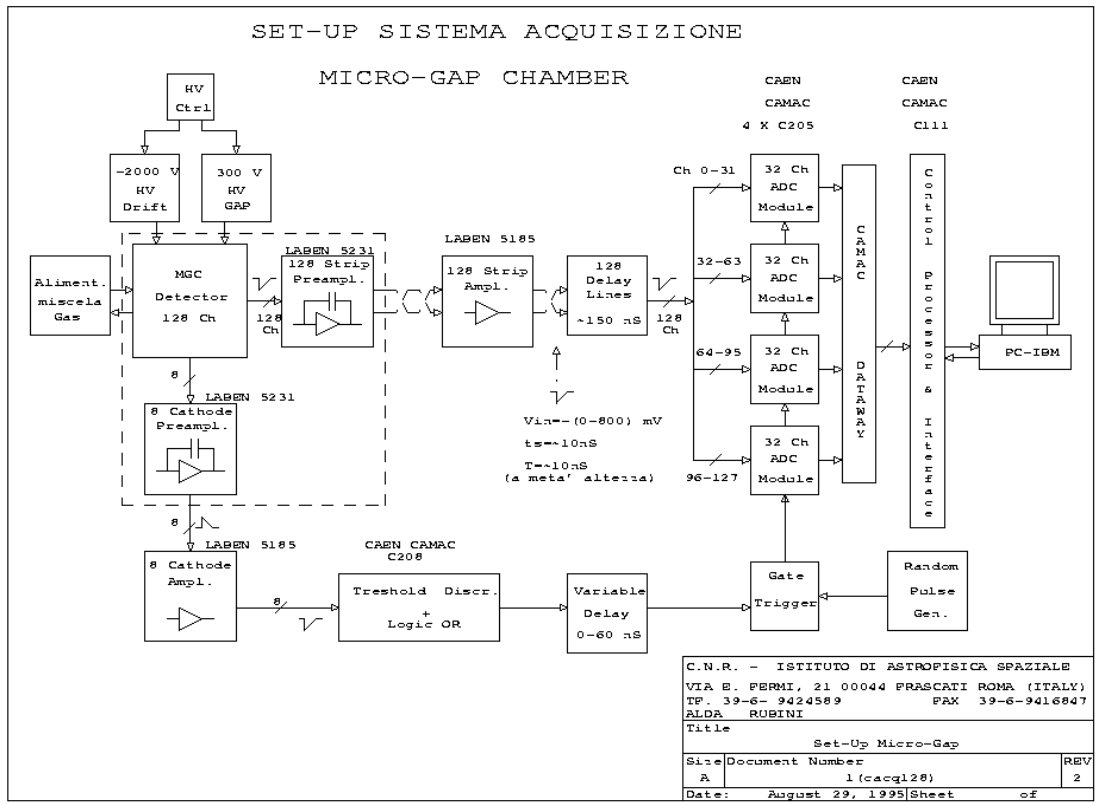

*Fig.2*

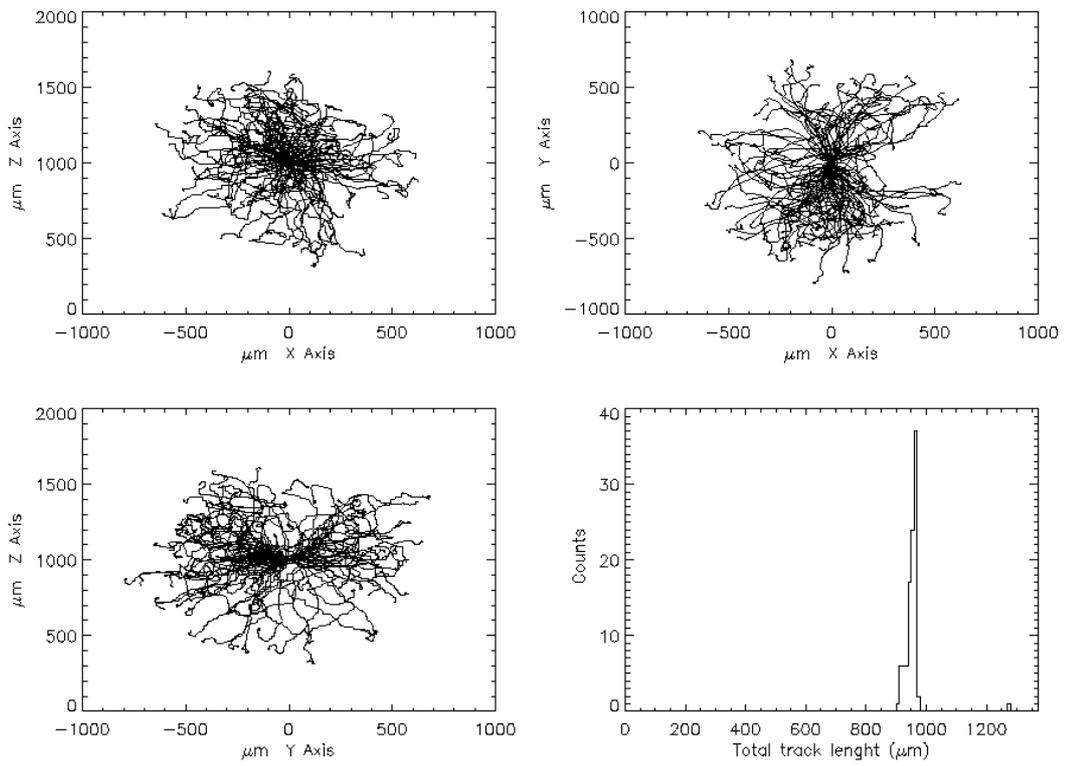

*Fig. 3*

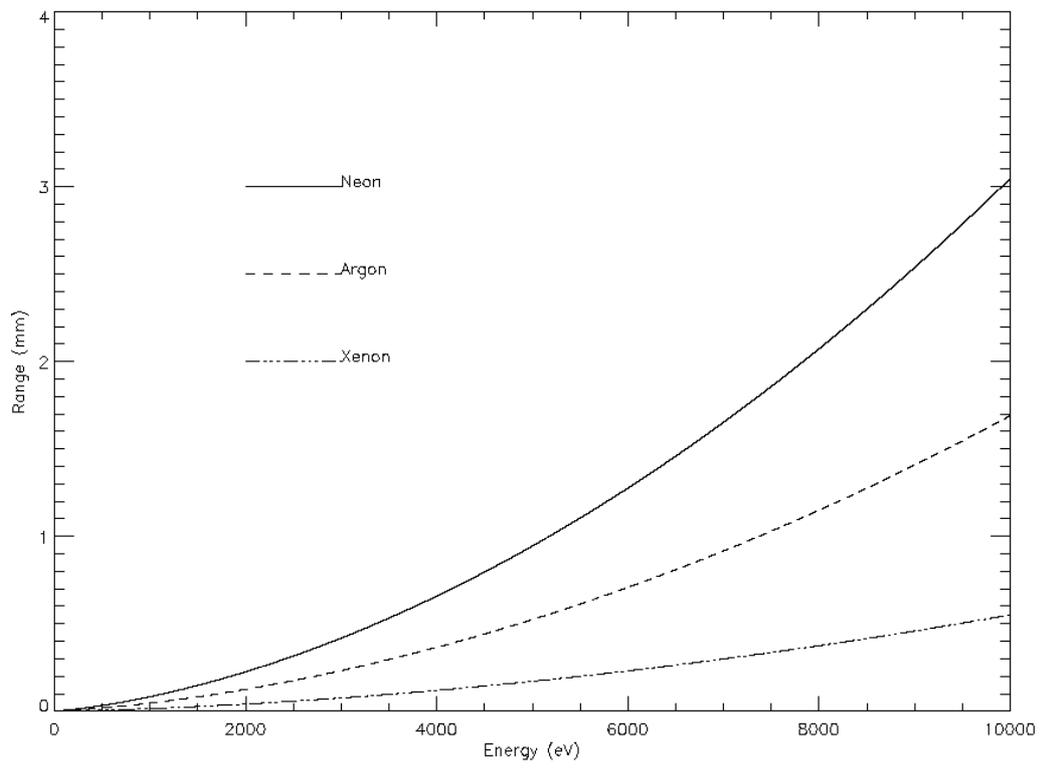

*Fig. 4*

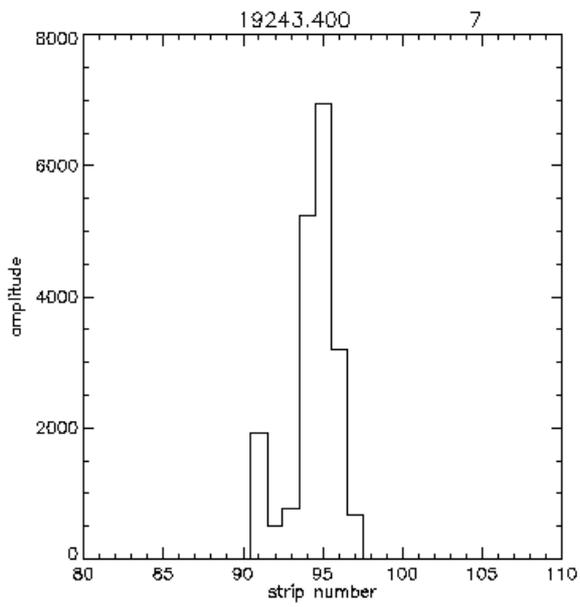
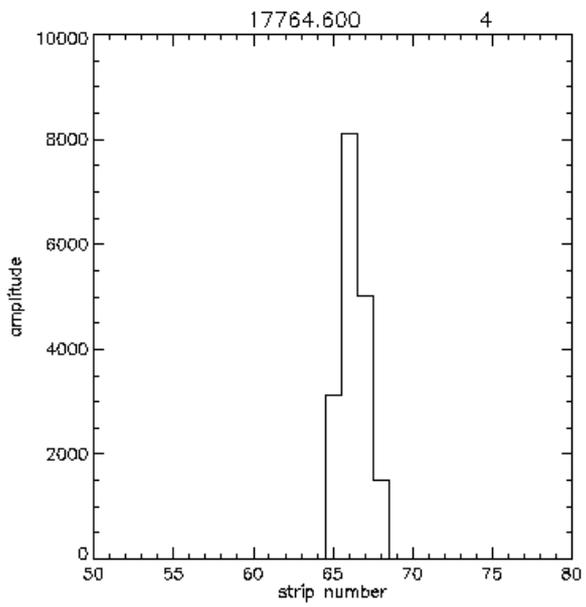
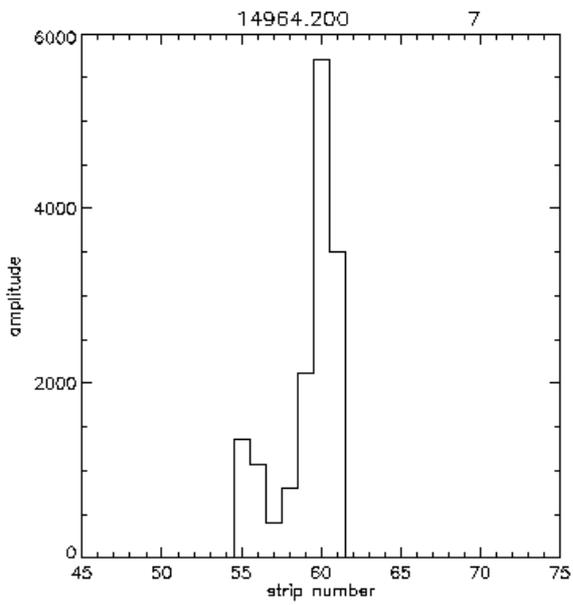
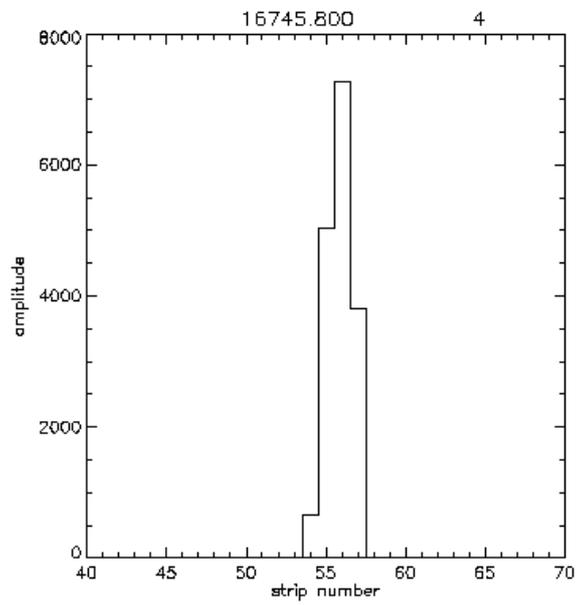

*Fig. 5*

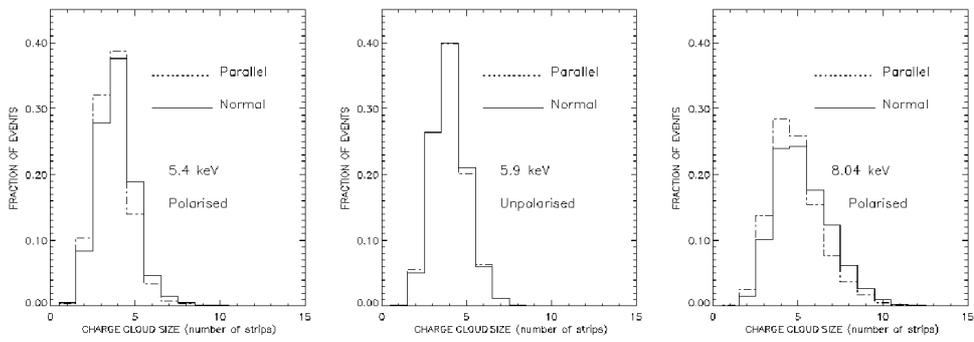

Fig. 6

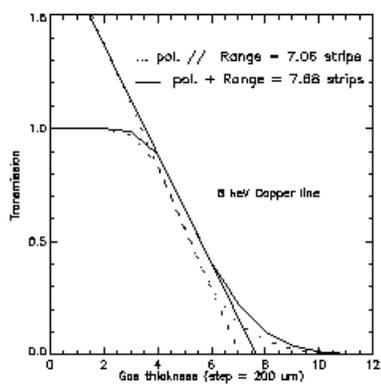 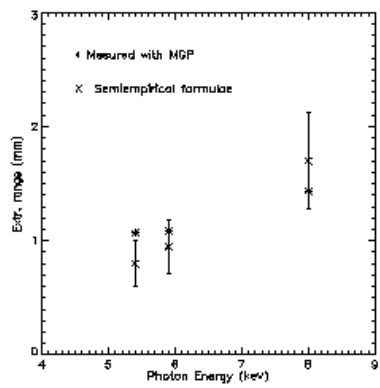

*Fig. 7*

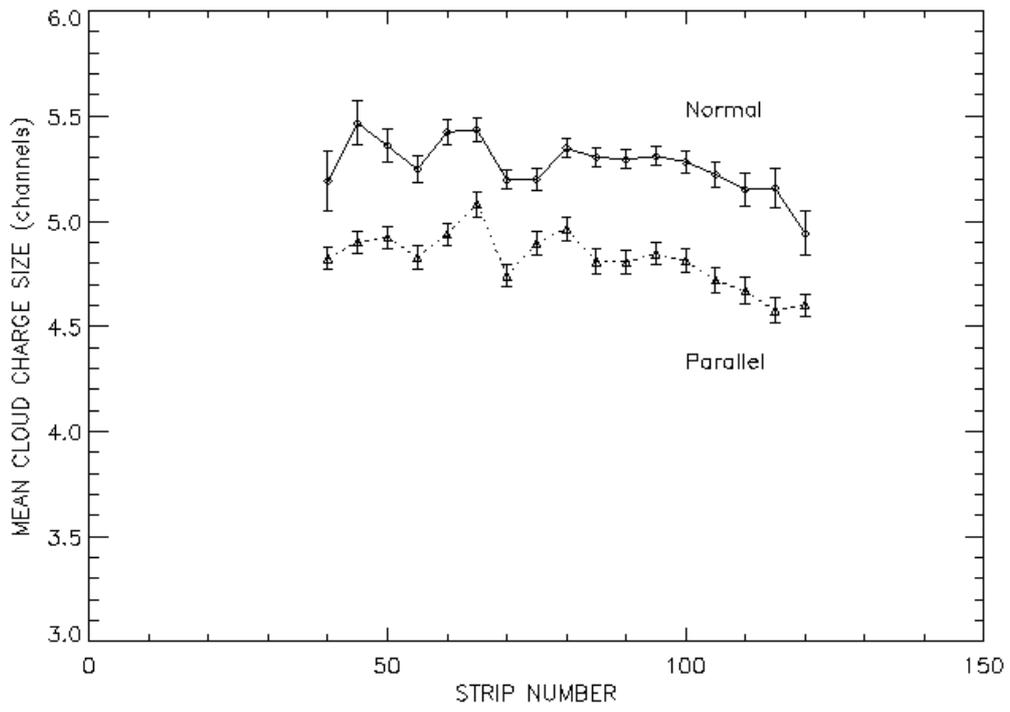

*Fig. 8*

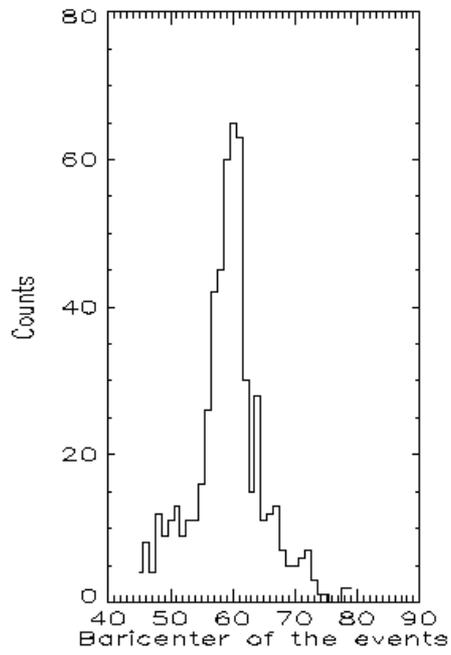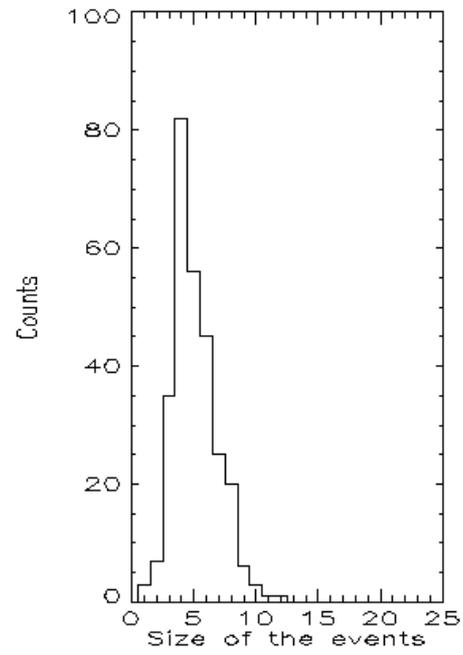
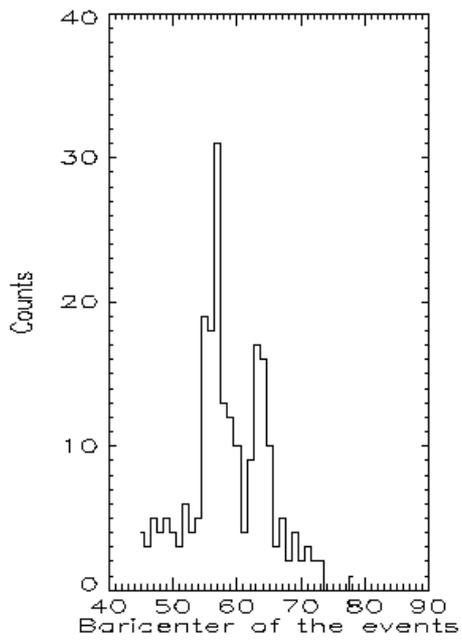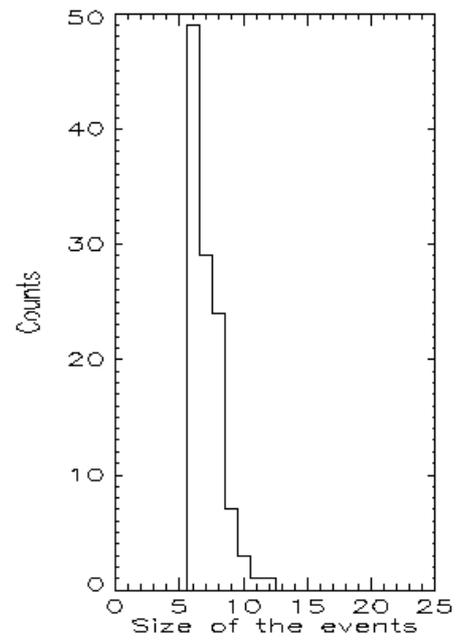

**Fig. 9**

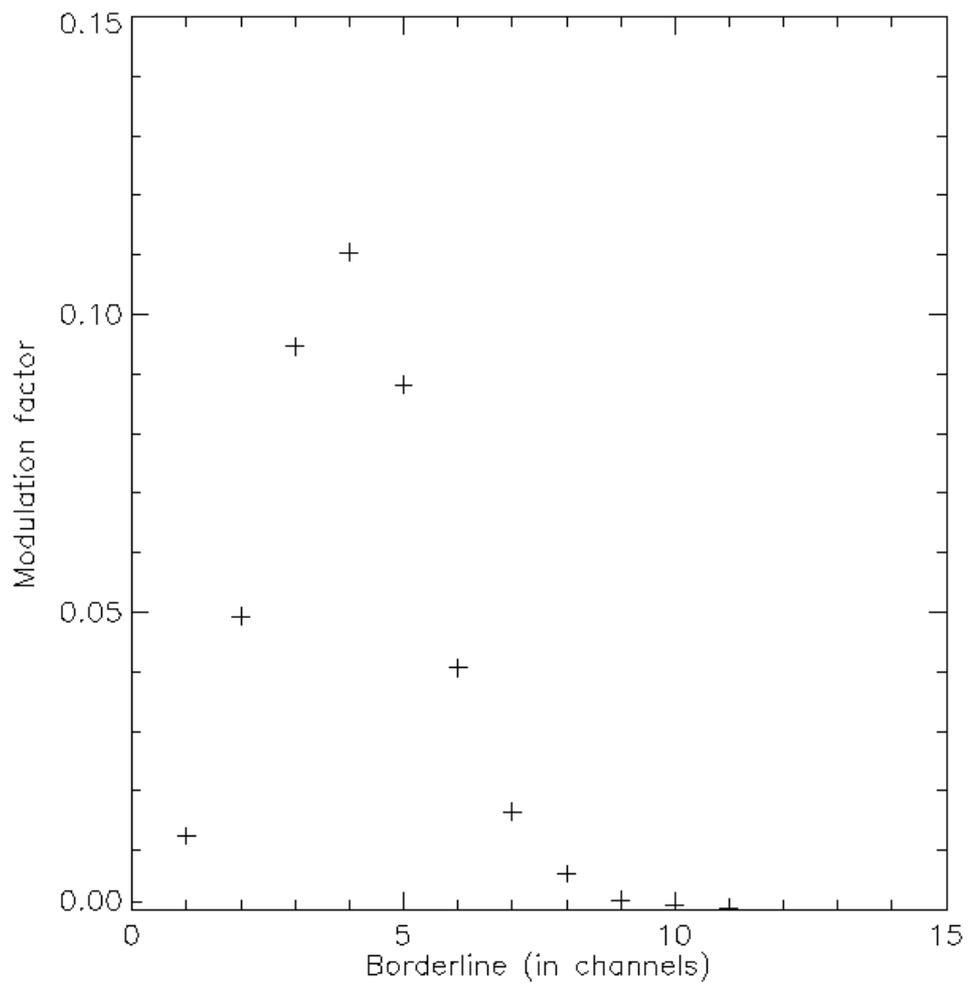

**Fig. 10**

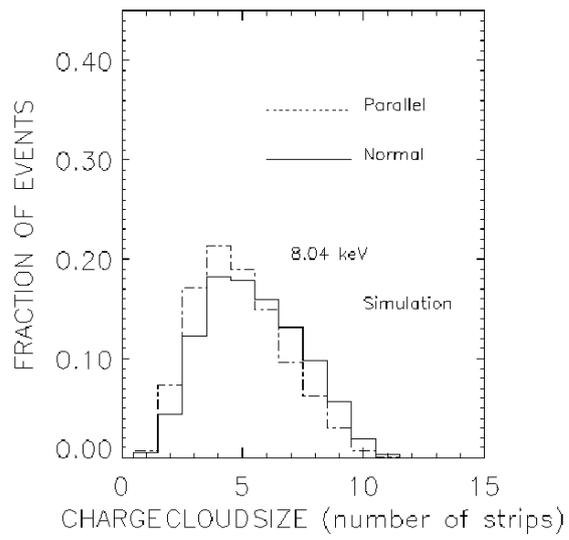

Fig. 11

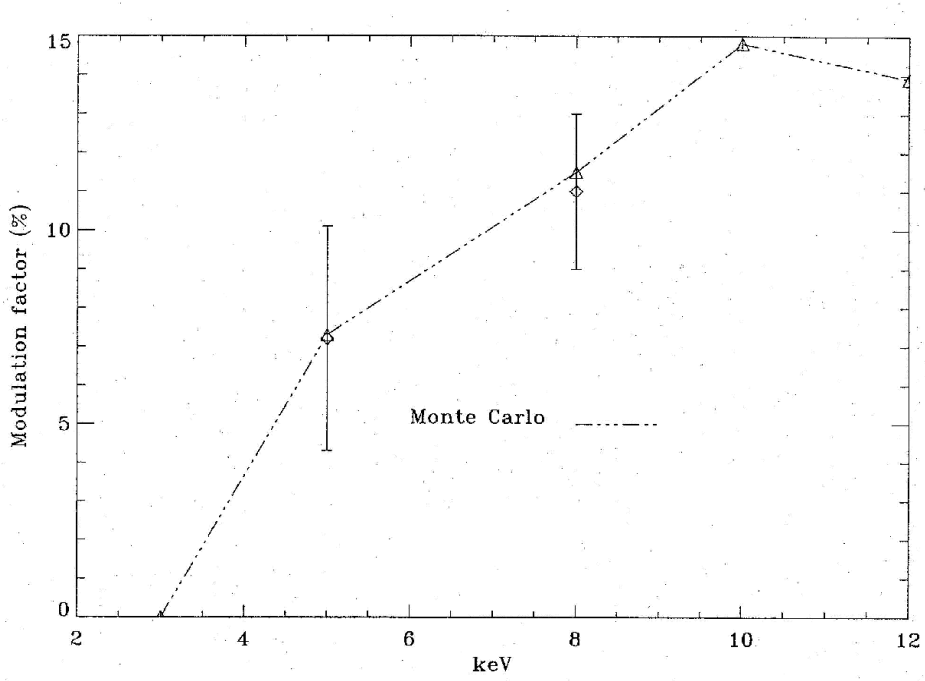

Fig. 12

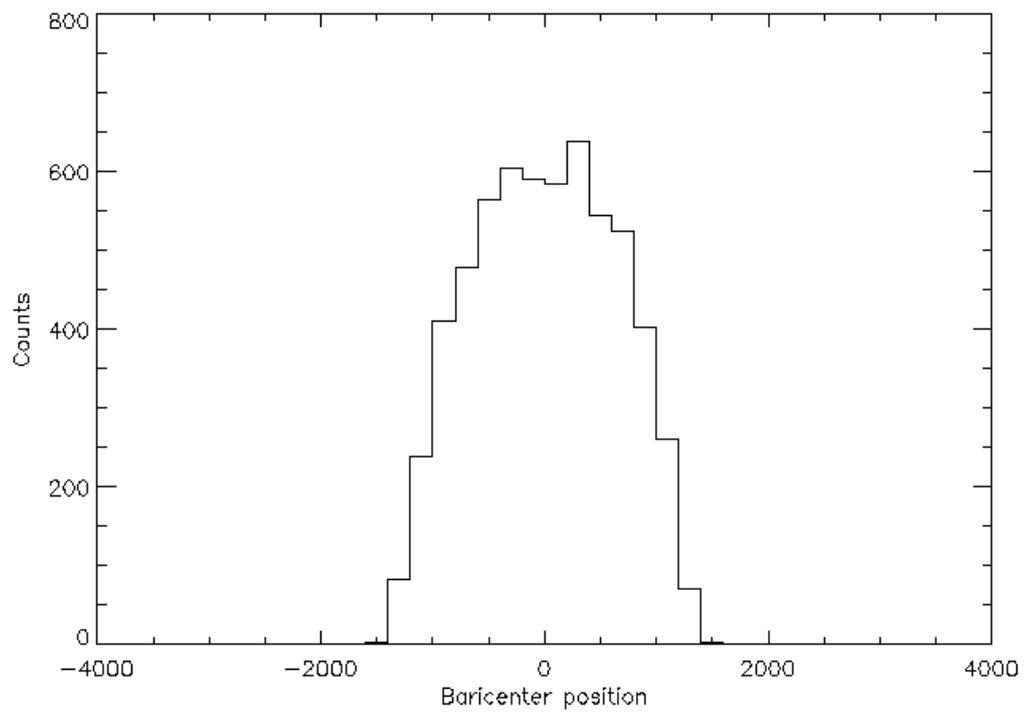

*Fig 13.1*

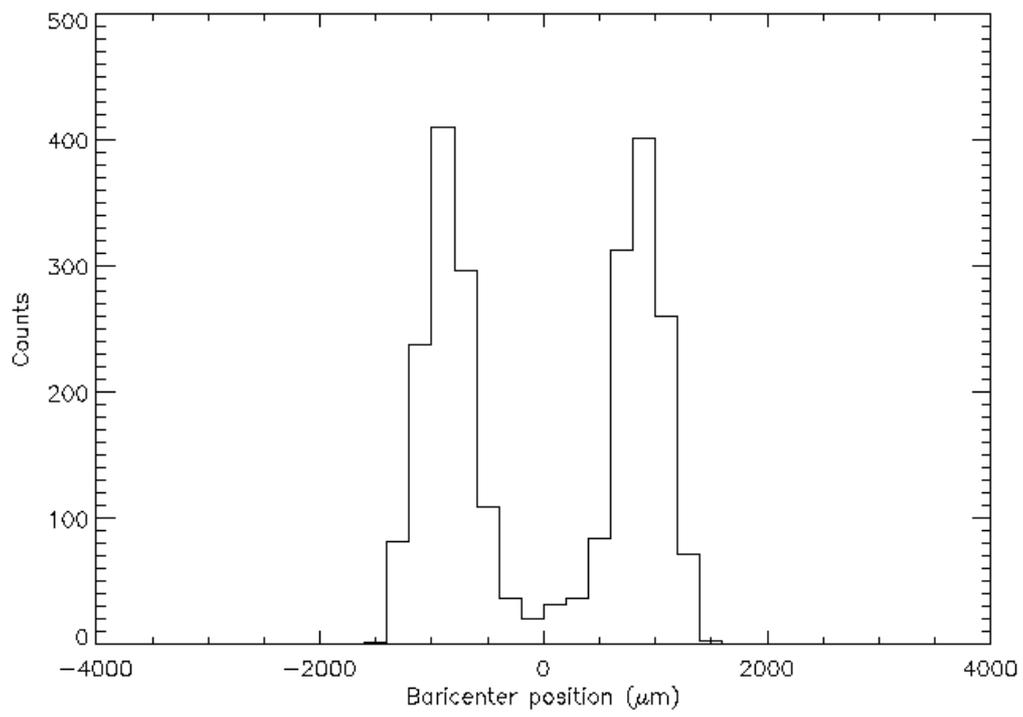

*Fig. 13.2*